\documentclass[10pt]{article}
\begin{document}
\pagestyle{plain}
\setcounter{page}{1}
\begin{center}
{\large\bf Perturbative Noncommutative Quantum Gravity}
\vskip 0.3 true in
{\large J. W. Moffat}
\vskip 0.3 true in
{\it Department of Physics, University of Toronto,
Toronto, Ontario M5S 1A7, Canada}
\end{center}
\begin{abstract}%
We study perturbative noncommutative quantum gravity by expanding the
gravitational field about a fixed classical background. A calculation of
the one loop gravitational self-energy graph reveals that only the
non-planar graviton loops are damped by oscillating internal momentum
dependent factors. The noncommutative quantum gravity perturbation theory
is not renormalizable beyond one loop for matter-free gravity and all loops
for matter interactions. Comments are made about the nonlocal gravitational
interactions produced by the noncommutative spacetime geometry.
\end{abstract}



\section{Introduction}

There has been renewed attention paid to noncommutative field theories in
view of their appearance in string theory and D-brane
theories~\cite{Connes,Witten,Douglas}. The noncommutative
geometry is characterized by $d$ noncommuting self-adjoint
operators ${\hat x}^\mu$ in a Hilbert space ${\cal H}$ satisfying
\begin{equation}
[{\hat x}^\mu,{\hat x}^\nu]=i\theta^{\mu\nu},
\end{equation}
where $\theta^{\mu\nu}$ is a non-degenerate $d\times d$ antisymmetric
matrix. Given an operator ${\hat\phi}$ associated with a
function $\phi(x)$ on the commutative space, the operator ${\hat\phi}$ acts
in the Hilbert space ${\cal H}$ according to the Weyl correspondence
\begin{equation}
{\hat\phi}({\hat x})=\frac{1}{(2\pi)^d}\int d^dxd^dk
\exp[ik_\mu({\hat x}^\mu-x^\mu)]\phi(x).
\end{equation}
The function $\phi(x)$ can be derived from
\begin{equation}
\phi(x)=\frac{1}{(2\pi)^{d/2}}\int d^dk\exp(ik_\mu x^\mu)
{\rm tr}[{\hat\phi}({\hat x})\exp(-ik_\mu{\hat x}^\mu)],
\end{equation}
where the trace operation ${\rm tr}$ is over the Hilbert space ${\cal H}$.

The product of two operators ${\hat\phi}_1$ and ${\hat\phi}_2$
has a corresponding Moyal $\star$-product
\begin{equation}
({\hat\phi}_1\star{\hat\phi}_2)(x)
=\exp\biggr(\frac{i}{2}\theta_{\mu\nu}\frac{\partial}{\partial\xi^\mu}\frac{\partial}
{\partial\zeta^\nu}\biggr)\phi_1(x+\xi)\phi_2(x+\zeta)\vert_{\xi=\zeta=0}.
\end{equation}

The problem with perturbative quantum gravity
based on a commutative spacetime and a local field theory
formalism is that the theory is not
renormalizable~\cite{Veltman,Van}. Due to the Gauss-Bonnet
theorem, it can be shown that the one loop graviton calculation
is renormalizable but two loop is not~\cite{Sagnotti}.
Moreover, gravity-matter interactions are not renormalizable at
any loop order.

Recently, the consequences for gravitation theory of using the Moyal
$\star$-product to define a general quantum gravity theory on a
noncommutative spacetime were analyzed, and it was found that a complex
symmetric metric (non-Hermitian) theory could possibly provide a consistent
underlying geometry on a complex coordinate manifold~\cite{Moffat}.
In the following, we shall investigate the consequences for
perturbative quantum gravity, when the gravitational action is given on a
noncommutative spacetime geometry. We expand the metric about a
flat Minkowski spacetime by taking the usual Einstein-Hilbert action, whose
fields are functions on ordinary noncommutative spacetime, except that the
products of field quantities are formed by using the Moyal $\star$-product
rule.

We treat $\theta^{\mu\nu}$ as an antisymmetric two-tensor with
$\theta_{\mu\nu}\theta^{\mu\nu}< 0$. The latter requirement guarantees that
the quantum theory is bounded from below in its lowest critical value. We
shall restrict ourselves to the case $\theta^{ij}\not=0$ and
$\theta^{0i}=0$, because the $\theta^{0i}\not=0$ case has peculiar causal
and quantum properties. The problem with time-space noncommutativity is due
to the fact that one cannot define the perturbative Hilbert space,
because the conjugate momentum of a field is ill-defined, and we cannot
know what the correct measure for the path integral should be. The
perturbative results for the gravitational field are derived by using the
modified graviton Feynman rules obtained by Filk~\cite{Filk}, and other
authors~\cite{Varilly,Martin,Chaichian,Ishibashi,Chepelev,Minwalla,Hawkins,Jabbari,Micu}.

The first order, one loop graviton self-energy is calculated in Sect.
3, using a noncommutative action and functional generator $Z[j_{\mu\nu}]$
~\cite{Feynman,Goldberg,Fradkin,Veltman,Leibbrandt,Brown,Medrano,Duff,Duff2,Donoghue}.
We find that the planar one loop graviton graph and vacuum polarization are
essentially the same as for the commutative perturbative result, while the non-planar
graviton loop graph is damped due to the oscillatory behavior of the noncommutative
phase factor in the Feynman integrand. Thus, the overall noncommutative perturbative
theory remains unrenormalizable and divergent. In Sect. 4, we consider the nonlocal
nature of the gravitional interactions in the noncommutative theory and in Sect 5, we
end with a summary of the results.

\section{\bf The Noncommutative Gravity Action}

We define the noncommutative gravitational action as
\begin{equation}
S_{\rm grav}=-\frac{2}{\kappa^2}\int d^4x(\sqrt{-g}\star R+2\sqrt{-g}\lambda),
\end{equation}
where we use the notation: $\mu,\nu=0,1,2,3$, $g={\rm det}(g_{\mu\nu})$, the
metric signature of Minkowski spacetime is $\eta_{\mu\nu}={\rm
diag}(-1,+1,+1,+1)$, $R=g^{\mu\nu}\star R_{\mu\nu}$ denotes the scalar curvature,
$\lambda$ is the cosmological constant and $\kappa^2=32\pi G$ with $c=1$. The Riemann
tensor is defined such that
\begin{equation}
{R^\lambda}_{\mu\nu\rho}=\partial_\rho{\Gamma_{\mu\nu}}^\lambda
-\partial_\nu{\Gamma_{\mu\rho}}^\lambda+
{\Gamma_{\mu\nu}}^\alpha\star{\Gamma_{\rho\alpha}}^\lambda
-{\Gamma_{\mu\rho}}^\alpha\star{\Gamma_{\nu\alpha}}^\lambda.
\end{equation}

We shall expand the gravity sector about flat Minkowski spacetime. In fact, we can
expand around any fixed, classical metric background~\cite{Veltman}
\begin{equation}
\label{background}
g_{\mu\nu}={\bar g}_{\mu\nu}+h_{\mu\nu},
\end{equation}
where ${\bar g}_{\mu\nu}$ is any smooth background metric field. For the sake of
simplicity, we shall only consider expansions about flat,
four-dimensional spacetime and we choose $\lambda=0$. Since the
gravitational field is weak up to the Planck energy scale, this expansion
is considered justified up to the latter scale; even at the standard model
energy scale $E_{\rm SM}\sim 10^2$ GeV, we have $\kappa E_{\rm SM}\sim
10^{-16}$. At these energy scales the curvature of spacetime is very small.

Let us define ${\bf g}^{\mu\nu}=\sqrt{-g}g^{\mu\nu}$.
It can be shown that $\sqrt{-g}=\sqrt{-{\bf g}}$, where ${\bf g}=
{\rm det}({\bf g}^{\mu\nu})$ and $\partial_\rho{\bf g}={\bf
g}_{\alpha\beta}\partial_\rho{\bf g}^{\alpha\beta}{\bf g}$.
We expand the local interpolating graviton field ${\bf g}^{\mu\nu}$ as
\begin{equation}
{\bf g}^{\mu\nu}=\eta^{\mu\nu}+\kappa\gamma^{\mu\nu}+O(\kappa^2).
\end{equation}
Then, for the noncommutative spacetime
\begin{equation}
{\bf g}_{\mu\nu}=\eta_{\mu\nu}-\kappa\gamma_{\mu\nu}
+\kappa^2{\gamma_\mu}^\alpha\star{\gamma_\alpha}_\nu-\kappa^3{\gamma_\mu}^\alpha
\star\gamma_{\alpha_\beta}\star{\gamma_\nu}^\beta+O(\kappa^4).
\end{equation}

We can now write the noncommutative gravitational action $S_{\rm grav}$ in
the form:
\begin{equation}
\label{action}
S_{\rm grav}=\frac{1}{2\kappa^2}\int
d^4x [({\bf g}^{\rho\sigma}\star{\bf g}_{\lambda\mu}\star{\bf g}_{\kappa\nu}
$$ $$
-\frac{1}{2}{\bf g}^{\rho\sigma}
\star{\bf g}_{\mu\kappa}\star{\bf g}_{\lambda\nu}
-2\delta^\sigma_\kappa\delta^\rho_\lambda{\bf
g}_{\mu\nu})\star\partial_\rho{\bf g}^{\mu\kappa}\star\partial_\sigma{\bf
g}^{\lambda\nu}].
\end{equation}

Since the free part of the action is the same as the commutative case, the vacuum
state and the quantization procedure are the same as in the commutative case. Only
the interaction part is linked to the noncommutative case. In particular, the measure
in the functional integral formalism is the same as the commutative theory, for in
momentum space the additional phase factor disappears when we impose the
normalization condition for the partition function.

Let us consider the noncommutative generating
functional~\cite{Fradkin,Leibbrandt,Medrano}
\begin{equation}
Z[j_{\mu\nu}]=\int
d[{\bf g}^{\mu\nu}]\Delta[{\bf g}^{\mu\nu}] \exp i\biggl[S_{\rm
grav}+\frac{1}{\kappa}\int d^4x{\bf g}^{\mu\nu}j_{\mu\nu} $$ $$
-\frac{1}{\kappa^2\alpha}\int d^4x\partial_\mu{\bf g}^{\mu\nu}
\star\partial_\alpha{\bf g}^{\alpha\beta}\eta_{\nu\beta}\biggr],
\end{equation}
where
$(\partial_\mu{\bf g}^{\mu\nu}\star\partial_\alpha{\bf
g}^{\alpha\beta}\eta_{\nu\beta}) /\kappa^2\alpha$ is the gauge fixing term.
Here, $\Delta$ can be interpreted in terms of fictitious particles and is
given by \begin{equation} \Delta[{\bf g}^{\mu\nu}]^{-1} =\int
d[\xi_\lambda]d[\eta_\nu]\exp i\biggl\{\int
d^4x\eta^\nu\star[\eta_{\nu\lambda}
\partial^\sigma\partial_\sigma-\kappa(\partial_\lambda\partial^\mu\gamma_{\mu\nu}-\gamma_{\mu\rho}
$$ $$
\eta_{\nu\lambda}\partial^\mu\partial^\rho-\partial^\mu\gamma_{\mu\rho}\eta_{\nu\lambda}
\partial^\rho+\partial^\mu\gamma_{\mu\nu}\partial_\lambda)]\star\xi^\lambda\biggr\},
\end{equation}
where $\xi^\lambda$ and $\eta^\lambda$ are the fictitious ghost particle fields.

The gravitational action is expanded as
\begin{equation}
S_{\rm grav}=S^{(0)}_{\rm grav}+\kappa S^{(1)}_{\rm grav}+\kappa^2
S^{(2)}_{\rm grav}+....
\end{equation}
We obtain
\begin{equation}
S^{(0)}_{\rm grav}=\frac{1}{2}\int d^4x[\partial_\sigma\gamma_{\lambda\rho}
\partial^\sigma\gamma^{\lambda\rho}-\partial_\lambda\gamma^{\rho\kappa}
\partial_\kappa\gamma^\lambda_\rho-\frac{1}{4}\partial_\rho\partial^\rho\gamma
-\frac{1}{\alpha}\partial_\rho\gamma^\rho_\lambda\partial_\kappa
\gamma^{\kappa\lambda}],
\end{equation}
\begin{equation}
S^{(1)}_{\rm grav}
=\int d^4x\frac{1}{4}[(-4\gamma_{\lambda\mu}\star\partial^\rho\gamma^{\mu\kappa}
\star\partial_\rho{\gamma_\kappa}^\lambda+2\gamma_{\mu\kappa}
\star\partial^\rho\gamma^{\mu\kappa}\star\partial_\rho\gamma
$$ $$
+2\gamma^{\rho\sigma}\star\partial_\rho\gamma_{\lambda\nu}
\star\partial_\sigma\gamma^{\lambda\nu}
-\gamma^{\rho\sigma}\star\partial_\rho\gamma\star\partial_\sigma\gamma
+4\gamma_{\mu\nu}\star\partial_\lambda\gamma^{\mu\kappa}
\star\partial_\kappa\gamma^{\nu\lambda})],
\end{equation}
\begin{equation}
S^{(2)}_{\rm grav}=\int d^4x[\frac{1}{4}(4\gamma_{\kappa\alpha}
\star\gamma^{\alpha\nu}
\star\partial^\rho\gamma^{\lambda\kappa}\star\partial_\rho\gamma_{\nu\lambda}
+(2\gamma_{\lambda\mu}\star\gamma_{\kappa\nu}-\gamma_{\mu\kappa}\star\gamma_{\nu\lambda})
\star\partial^\rho\gamma^{\mu\kappa}\star\partial_\rho\gamma^{\nu\lambda}
$$ $$
-2\gamma_{\lambda\alpha}\star{\gamma_\nu}^\alpha\star\partial^\rho\gamma^{\lambda\nu}
\star\partial_\rho\gamma-2\gamma^{\rho\sigma}\star{\gamma_\nu}^\kappa
\star\partial_\rho\gamma_{\lambda\kappa}\star\partial_\sigma\gamma^{\nu\lambda}
$$ $$
+\gamma^{\rho\sigma}\star\gamma^{\nu\lambda}\star\partial_\sigma\gamma_{\nu\lambda}
\star\partial_\rho\gamma-2\gamma_{\mu\alpha}\star\gamma^{\alpha\nu}
\star\partial^\lambda\gamma^{\mu\kappa}\star\partial_\kappa\gamma_{\nu\lambda})],
\end{equation}
where $\gamma={\gamma^\alpha}_\alpha$. We have taken into account that the kinetic
term in the action $S^{(0)}_{\rm grav}$ involves quadratic $\star$-product terms
under the integral sign and is the same as the commutative case. Thus, for example,
\begin{equation}
\int d^4x\partial_\sigma\gamma_{\lambda\rho}
\star\partial^\sigma\gamma^{\lambda\rho} $$ $$ =\int
d^4x\partial_\sigma\gamma_{\lambda\rho} \partial^\sigma\gamma^{\lambda\rho}.
\end{equation}
It follows from this that the two-point graviton propagator is the same as in the
commutative theory. The graviton propagator in the fixed de
Donder gauge $\alpha=-1$~\cite{Donder} is given by
\begin{equation}
D^{\rm
grav}_{\mu\nu\rho\sigma}(x)
=(\eta_{\mu\rho}\eta_{\nu\sigma}+\eta_{\mu\sigma}\eta_{\nu\rho}
-\eta_{\mu\nu}\eta_{\rho\sigma}) $$ $$
\times\biggl(\frac{-i}{(2\pi)^4}\biggr)\int
\frac{d^4p}{p^2-i\epsilon}\exp[ip\cdot(x-x')].
\end{equation} In momentum
space this becomes
\begin{equation}
D^{\rm grav}_{\mu\nu\rho\sigma}(p)
=(\eta_{\mu\rho}\eta_{\nu\sigma}+\eta_{\mu\sigma}\eta_{\nu\rho}
-\eta_{\mu\nu}\eta_{\rho\sigma})\frac{1}{p^2}.
\end{equation}
The ghost propagator in momentum space is given by
\begin{equation}
D^G_{\mu\nu}(p)=\frac{\eta_{\mu\nu}}{p^2}.
\end{equation}
 
Let us consider the effects of an infinitesimal gauge transformation
\begin{equation}
x^{'\mu}=x^\mu+\zeta^\mu
\end{equation}
on the noncommutative generating functional $Z[j_{\mu\nu}]$, where $\zeta^\mu$ can
depend on $x^\mu$ and $\gamma^{\mu\nu}$. We get
\begin{equation}
\delta{\bf g}^{\mu\nu}(x)
=-\zeta^\lambda(x)\star\partial_\lambda{\bf g}^{\mu\nu}(x)
+\partial_\rho\zeta^\mu(x)\star{\bf g}^{\rho\nu}(x) $$ $$
+\partial_\sigma\zeta^\nu(x)\star{\bf g}^{\mu\sigma}(x)
-\partial_\alpha\zeta^\alpha(x)\star{\bf g}^{\mu\nu}(x). \end{equation} We now find
that
\begin{equation}
\label{gaugetransf}
\delta\gamma_{\mu\nu}(x)=-\zeta^\lambda\star\partial_\lambda\gamma_{\mu\nu}
+\partial^\rho\zeta_\mu\star\gamma_{\rho\nu}+\partial^\rho\zeta_\nu\star\gamma_{\mu\rho}
$$ $$ -\partial^\rho\zeta_\rho\star\gamma_{\mu\nu}
+\frac{1}{\kappa}(\partial_\nu\zeta_\mu+\partial_\mu\zeta_\nu-\partial^\rho\zeta_\rho\eta_{\mu\nu}).
\end{equation}
The functional generator $Z$ is invariant under changes in the
integration variable and the transformation $(\ref{gaugetransf})$.

The noncommutative gauge transformations (\ref{gaugetransf}) should be
considered part of an $NCSO(3,1)$ group of gauge transformations, which
could correspond in general to a complex gravity theory of the kind
discussed in refs.~\cite{Chamseddine,Moffat}. Bonora et al.~\cite{Bonora}
have shown that local gauge theories in commutative spaces with $\theta=0$
do not trivially extend themselves to noncommutative spaces. It is clear
that in the limit $\theta\rightarrow 0$ the standard local Lorentz group of
gauge transformations $SO(3,1)$ is recovered. By considering charge
conjugation operation in gauge theories, Bonora et al. can construct a
$NCSO(3,1)$ group of gauge transformations.

\section{Gravitational Self-Energy}

The lowest order contributions to the graviton self-energy will include
the standard graviton loops, the ghost field loop contributions and the measure loop
contributions. In perturbative gravity theory, the first order vacuum
polarization tensor $\Pi^{\mu\nu\rho\sigma}$ must satisfy the Slavnov-Ward
identities~\cite{Medrano}:
\begin{equation}
\label{SlavnovWard}
p_\mu p_\rho
D^{\mu\nu\alpha\beta}(p)\Pi_{\alpha\beta\gamma\delta}(p)
D^{\gamma\delta\rho\sigma}(p)=0.
\end{equation}
By symmetry and Lorentz invariance, the vacuum polarization tensor must have the form
\begin{equation}
\Pi_{\alpha\beta\gamma\delta}(p)
=\Pi_1(p^2)p^4\eta_{\alpha\beta}\eta_{\gamma\delta}+\Pi_2
(p^2)p^4(\eta_{\alpha\gamma}\eta_{\beta\delta}
+\eta_{\alpha\delta}\eta_{\beta\gamma}) $$ $$
+\Pi_3(p^2)p^2(\eta_{\alpha\beta}p_\gamma p_\delta+\eta_{\gamma\delta}p_\alpha
p_\beta) +\Pi_4(p^2)p^2(\eta_{\alpha\gamma}p_\beta
p_\delta+\eta_{\alpha\delta}p_\beta p_\gamma $$ $$ +\eta_{\beta\gamma}p_\alpha
p_\delta+\eta_{\beta\delta}p_\alpha p_\gamma)+\Pi_5(p^2)p_\alpha p_\beta p_\gamma
p_\delta.
\end{equation}
The Slavnov-Ward identities impose the restrictions
\begin{equation}
\Pi_2+\Pi_4=0,\quad 4(\Pi_1+\Pi_2-\Pi_3)+\Pi_5=0.
\end{equation}

For noncommutative gravity theory, it is useful to treat the
metric $g_{\mu\nu}$ as an $N\times N$ matrix and to employ double line
notation when calculating Feynman graphs. The planar graphs and the
non-planar graphs are not equivalent for $\theta\not=0$. For the planar
graviton graphs, we can treat the momentum as an additional index flowing
along double lines~\cite{Hooft}. For an L loop planar graph, the momenta of
all lines may be expressed in terms of momenta $r_1,...r_{L+1}$ that run
along an index line of the graph. The momentum through any propagator or
external line is given by $r_i-r_j$, where $r_i$ and $r_j$ are the index
momenta that run along the sides of the propagator. This configuration
automatically guarantees momentum conservation at the vertices. The index
momenta along adjacent sides of a graviton propagator move in opposite
directions. This construction is not valid for non-planar graphs.

In momentum space, a graviton interaction vertex has an additional phase
factor relative to the commutative
theory~\cite{Filk,Chepelev,Minwalla}
\begin{equation}
V(q_1,q_2...,q_n)=\exp\biggl(-\frac{i}{2}\sum_{i<j}q_i\times q_j\biggr),
\end{equation}
where
\begin{equation}
q_i\times q_j\equiv q_{i\mu}\theta^{\mu\nu}q_{j\nu}.
\end{equation}
In flat spacetime, this is the only change of the Feynman rules.
Using momentum conservation, we find that $V(q_1,...,q_n)$ is invariant
under the cyclic permutations of $q_i$.

For a planar graph (genus zero), the phase factor associated with any
internal propagator is equal and opposite at its two end vertices, thereby
cancelling identically. Thus, the planar graviton graph has a phase factor
\begin{equation}
V(p_1,p_2,...,p_n)=\exp\biggl(-\frac{i}{2}\sum_{i<j}p_i\times p_j\biggr),
\end{equation}
where the sum is taken over all external momenta in the appropriate cylic
order.

The basic lowest order graviton self-energy diagram is determined by
\begin{equation}
\Pi^{\rm planar}_{\mu\nu\rho\sigma}(p)
=\frac{1}{2}\kappa^2\exp\biggl(-\frac{i}{2}\sum_{i<j}p_i\times
p_j\biggr)\int d^4q {\cal U}_{\mu\nu\alpha\beta\gamma\delta}(p,-q,q-p)
D^{{\rm grav}\alpha\beta\kappa\lambda}(q)
$$ $$
\times D^{{\rm grav}\gamma\delta\tau\xi}(p-q){\cal
U}_{\kappa\lambda\tau\xi\rho\sigma}(q,p-q,-p),
\end{equation}
where $i,j=1,2,3$ and ${\cal U}$ is the three-graviton vertex function
\begin{equation}
{\cal U}_{\mu\nu\rho\sigma\delta\tau}(q_1,q_2,q_3) =
-\frac{1}{2}\biggl[q_{2(\mu}q_{3\nu)}\biggl(2\eta_{\rho(\delta}\eta_{\tau)\sigma}
-\eta_{\rho\sigma}\eta_{\delta\tau}\biggr)
$$ $$
+q_{1(\rho}q_{3\sigma)}\biggl(2\eta_{\mu(\delta}\eta_{\tau)\nu}
-\eta_{\mu\nu}\eta_{\delta\tau}\biggr)+...\biggr],
\end{equation}
and the ellipsis denote similar contributions. The sum in the phase factor
is taken over all the external momenta $p_i$. We must add to this result
the contributions from the planar one loop fictitious ghost particle graph
and tadpole graph, in which the noncommutative phase factor
again only depends on the external momenta. The $\theta$ dependent phase
factor is present in all graviton interaction terms and in all graviton
tree planar graphs.

The non-planar graviton graphs have propagators that intersect one another
or cross external lines. Any non-planar graph has an additional phase
$\exp(iq_i\times q_j)$ for each momenta $q_i$ and $q_j$ that cross, in
addition to the external momenta phase factor. The complete
phase for a general graviton graph is
\begin{equation}
W(p_1,p_2,...p_n;q_1,q_2,...q_n)=V(p_1,p_2,...,p_n)
\exp\biggl[-i\biggl(\frac{1}{2}\sum_{i,j}I_{ij}q_i\times q_j\biggr)\biggr],
\end{equation}
where $I_{ij}$ is the intersection matrix that counts the number of times
the ith graviton line crosses over the jth graviton line.

The non-planar one loop self-energy diagram is given by
\begin{equation}
\Pi^{\rm nonplanar}_{\mu\nu\rho\sigma}(p)
=\frac{1}{2}\kappa^2\exp\biggl(-\frac{i}{2}\sum_{i<j}p_i\times
p_j\biggr)\int d^4q \exp(iq_i\times p_j)
{\cal U}_{\mu\nu\alpha\beta\gamma\delta}(p,-q,q-p)
$$ $$
D^{{\rm grav}\alpha\beta\kappa\lambda}(q)
D^{{\rm grav}\gamma\delta\tau\xi}(p-q){\cal
U}_{\kappa\lambda\tau\xi\rho\sigma}(q,p-q,-p).
\end{equation}
To this diagram, we must add the non-planar fictitious
ghost particle graph and tadpole graph contributions, which also have an
additional internal momentum phase factor. The final graviton self-energy
one loop, non-planar result for the noncommutative theory will be
convergent for values of the external momenta, because of the rapid
oscillations of the phase factor $\exp(iq\times p)$, where $q$ is an
internal momentum and $p$ is an external momentum. However, the non-planar
graph is singular when $\vert p_\mu\theta^{\mu\nu}\vert$ vanishes, because
the phase factor is then zero. The effective gravitational cutoff in
momentum space is
\begin{equation}
\Lambda_{\rm grav}=\frac{1}{\sqrt{-p_\mu\theta^2_{\mu\nu}p_\nu}}.
\end{equation}
Therefore, as in scalar field theories~\cite{Minwalla}, the perturbative
quantum gravity theory has a peculiar mixture of ultraviolet and infrared
divergent behavior. Switching on $\theta$ can replace the ultraviolet
divergence with a singular infrared behavior.

We see that the divergence of the planar one loop self-energy graph is
essentially the same as for the commutative perturbative result.
This basic result will hold for higher order loops, as well. Therefore, we
conclude that, except for the first order loop graph in matter-free
gravity, noncommutative perturbative quantum gravity is unrenormalizable.
However, the situation with noncommutative renormalizability of field theories that
are renormalizable in the $\theta=0$ commutative limit is not
straightforward~\cite{Chepelev,Minwalla}. For scalar noncommutative
quantum field theory, the class of renormalizable, topologically nontrivial
diagrams is smaller than the class of all diagrams in the theory, which
leaves open the question of perturbative renormalizability of such
noncommutative field theories.

In the framework of an effective gravitational field
theory~\cite{Donoghue}, the leading lowest order loop divergence can be
``renormalized'' by being absorbed into two parameters $c_1$ and $c_2$. For
a non-flat spacetime background metric ${\bar g}_{\mu\nu}$, the divergent
noncommutative term at one loop due to graviton and ghost loops is given in the
Lagrangian by~\cite{Veltman}:
\begin{equation}
{\cal L}^{\rm div}_{1{\rm
loop}}=\frac{1}{8\pi^2\epsilon} \biggl[\frac{1}{120}{\bar R}\star{\bar
R}+\frac{7}{20}{\bar R}_{\mu\nu} \star{\bar R}^{\mu\nu}\biggr], \end{equation} where
in the dimensional regularization method $\epsilon=4-d$, and the effective 
field theory renormalization parameters are
\begin{equation}
c^{(r)}_1=c_1+\frac{1}{960\pi^2\epsilon},\quad
c^{(r)}_2=c_2+\frac{7}{160\pi^2\epsilon}.
\end{equation}
 
\section{Nonlocality and Unitarity}

The infinite derivatives that occur in the Moyal $\star$-product render the
noncommutative field theories nonlocal. This is certainly true for
noncommutative quantum gravity. If we write the star product of two first
order $\gamma_{\mu\nu}$ in position space as~\cite{Minwalla}:
\begin{equation}
(\gamma_{1\mu\nu}\star\gamma_{2\sigma\rho})(z)=
\int d^4z_1d^4z_2\gamma_{1\mu\nu}(z_1)\gamma_{2\sigma\rho}(z_2)K(z_1,z_2,z),
\end{equation}
where the kernel $K$ is given by
\begin{equation}
K(z_1,z_2,z)=\frac{1}{{\rm
det\theta}}\exp[2i(z-z_1)^\mu\theta_{\mu\nu}^{-1}(z-z_2)^\nu],
\end{equation}
then in view of the fact that $\vert K\vert$ is a constant independent of
$z_1,z_2$ and $z$, the $\star$-product appears to be infinitely
nonlocal, although the oscillations in the phase of $K$ damp out parts of
the integration. If $\gamma_{\mu\nu}$ is nonzero over a tiny region of
space of size $\Delta \ll \sqrt{\theta}$, then $\gamma\star\gamma$ is
non-vanishing over a much larger region of size $\theta/\Delta$.

This nonlocal behaviour of the gravitational interactions has serious
consequences for the dynamics. In particular, if $\theta^{i0}\not=0$, then
there is a nonlocality in time which can violate unitarity of the graviton
scattering amplitudes~\cite{Gomis} and disqualify the use of standard
quantum field theory techniques, such as canonical Hamiltonian
quantization. These problems surface in the investigation of open
string theory calculations of scattering
amplitudes~\cite{Seiberg,Seiberg2}. It is far from obvious why nature
should choose to retain only the $\theta^{ij}\not=0$ structure of the
noncommutative theory, and not include the more problematical
$\theta^{0i}\not=0$ structure. Such a circumstance would render the whole
scheme non-covariant and mathematically, if not physically, displeasing.

One important problem that can arise, because
of the nonlocal nature of the noncommutative field theories is that
the theories can exhibit instability problems. This has
always been an open question in string theories and in D-brane
physics~\cite{Eliezer,Kamimura,Douglas}. It is certainly also an open question in
our perturbative noncommutative quantum gravity.

\section{\bf Conclusions}

We have developed a perturbative, noncommutative quantum gravity formalism
by using the Moyal $\star$-product in the gravity action, wherever products
of gravitational fields and their derivatives occur. By expanding about
Minkowski flat spacetime, we were able to calculate the planar and
non-planar loop graphs to first order. We find that by using the Feynman
rules appropriate for noncommutative quantum field theory, the planar
loop graph was essentially the same as in commutative perturbative quantum
gravity, up to a non-zero phase factor that only depends on the external
momenta. Only the non-planar loops exhibit convergence properties,
because of the non-vanishing phase factors that depend on the internal loop
momenta, and this was due to their generically oscillatory behavior. The
dependence on the cut-off $\Lambda_{\rm grav}$ was such that there is a
peculiar mixture of ultraviolet and infrared singular behavior.

The main disappointing result of our investigation is that perturbative
noncommutative quantum gravity is not renormalizable, due to the divergent
behavior of all the planar loop graphs to all orders, giving rise to new
coefficients in the graviton self-energy contributions at every order
and, therefore, yielding an infinite number of parameters to all orders. This
result should perhaps come as a surprise, because the assumption that the
spacetime coordinates are noncommuting is quite a drastic one; it
implies the existence of an essentially lattice structure of spacetime.
Due to the ``fuzzy'' nature of spacetime that arises from the Heizenberg
uncertainty principle
\begin{equation}
\Delta x^\mu\Delta x^\nu \ge \vert\theta^{\mu\nu}\vert,
\end{equation}
and the intrinsic nonlocality of the interactions, we might expect that the
quantum gravity calculations would be finite to all orders. Of course,
these results are based on perturbative calculations, which do not
provide an answer to whether a non-perturbative, noncommutative quantum
gravity formalism could produce a finite, unitary and gauge invariant
graviton scattering amplitude. We are far from being able answer this
question in the affirmative.

In ref. [11], a complex symmetric metric formulation of quantum
gravity was investigated, in which a Moyal $\diamond$-product of functions 
was introduced for {\it anticommuting} coordinates. This product was 
defined in terms of a symmetric two-tensor $\tau^{\mu\nu}=\tau^{\nu\mu}$ 
and could lead to possible convergence of planar loop graphs.

The nonlocal nature of the noncommutative
gravitational interactions, raises some serious questions about the
physical stability of such a theory and whether the whole scheme leads to
physically acceptable consequences. Clearly, further investigation of these
questions is needed before we can accept that a noncommutative field
theory program, and a noncommutative quantum gravity theory, can lead to
physically viable theories of nature. All of these issues do have
important ramifications for string theory and recent developments in
D-brane physics.

\vskip 0.2
true in {\bf Acknowledgments}
\vskip 0.2 true in
This work was supported by the Natural Sciences and Engineering Research Council of
Canada.
\vskip 0.5 true in


\begin{thebibliography}{100}

\bibitem{Connes} A. Connes, M. R. Douglas and A. Schwarz, JHEP 9802 (1998)
003, hep-th/9711162.

\bibitem{Witten} N. Seiberg and E. Witten, JHEP 9909 (1999) 032,
hep-th/9908142.

\bibitem{Douglas} For a review and references, see: M. R.
Douglas, ``Two Lectures on D-Geometry and Noncommutative Geometry",
hep-th/9901146.

\bibitem{Veltman} G. 't Hooft and M. Veltman, Ann. Inst. Henri
Poincar\'e, {\bf 30}, 69 (1974).

\bibitem{Van} S. Deser and P. van Nieuwenhuizen, Phys. Rev.
{\bf 10}, 401 (1974); Phys. Rev. {\bf 10}, 411 (1974).
 
\bibitem{Sagnotti} M. Goroff and A. Sagnotti, Nucl. Phys. {\bf
B266}, 709 (1986).

\bibitem{Moffat} J. W. Moffat, hep-th/0007181. To be published in Phys.
Lett. B.

\bibitem{Filk} T. Filk, Phys. Lett. {\bf B376}, 53 (1996).

\bibitem{Varilly} J. C. Varilly and J. M. Gracia-Bondia, Int. J. Mod. Phys.
{\bf A14}, 1305 (1999), hep-th/9804001.

\bibitem{Martin} C. P. Martin and D. Sanchez-Ruiz, Phys. Rev. Lett. {\bf
83}, 476 (1999), hep-th/9903077.

\bibitem{Chaichian} M. Chaichian, A. Demichev and P. Prenajder, Nucl. Phys.
B{\bf 567}, 360 (2000), hep-th/9812180.

\bibitem{Ishibashi} N. Ishibasi, S. Iso, H. Kawai and Y. Kitazawa,
hep-th/9910004.

\bibitem{Chepelev} I. Chepelev and R. Roiban, JHEP 0005 (2000) 037,
hep-th/9911098 v4.

\bibitem{Minwalla} S. Minwalla, M. Van Raamsdonk and N. Seiberg,
hep-th/9912072 v2; M. Raamsdonk and N. Seiberg, JHEP 0003 (2000) 035,
hep-th/0002186.

\bibitem{Hawkins} E. Hawkins, hep-th/9908052.

\bibitem{Jabbari} M. M. Sheikh-Jabbari, JHEP 9906 (1999) 015,
hep-th/9903107.

\bibitem{Micu} A. Micu and M. M. Sheikh-Jabbari, hep-th/0008057.

\bibitem{Feynman} R. P. Feynman, Acta Phys.
Pol. {\bf 24}, 697 (1963); Magic Without Magic, edited by J.
Klauder (Freeman, New York, 1972), p. 355; Feynman Lectures on
Gravitation, edited by B. Hatfield, (Addison-Wesley publishing
Co. 1995.)

\bibitem{Goldberg} J. N. Goldberg, Phys. Rev. {\bf 111}, 315
(1958).
 
\bibitem{Fradkin} E. S. Fradkin and I. V. Tyutin, Phys. Rev.
{\bf D2}, 2841 (1970).
 
\bibitem{Leibbrandt} D. M. Capper, G. Leibbrandt, and M. R.
Medrano, Phys. Rev. {\bf D8}, 4320 (1973).

\bibitem{Brown} M. R. Brown, Nucl. Phys. {\bf B56}, 194 (1973).

\bibitem{Medrano}  D. M. Capper and M. R. Medrano, Phys. Rev.
{\bf D9}, 1641 (1974).
 
\bibitem{Duff} D. M. Capper, M. J. Duff, and L. Halpern, Phys.
Rev. {\bf D10}, 461 (1974).
 
\bibitem{Duff2} M. J. Duff, Phys. Rev. {\bf D9}, 1837 (1974).

\bibitem{Donoghue} J. F. Donoghue, Phys. Rev. {\bf D50}, 3874 (1994).

\bibitem{Chamseddine} A. H. Chamseddine, hep-th/0005222.

\bibitem{Bonora} L. Bonora, M. Schnabl, M. M. Sheikh-Jabbari and A.
Tomasiello, hep-th/0006091; M. M. Sheikh-Jabbari, hep-th/0001167.

\bibitem{Donder} T. de Donder, {\it La Grafique Einsteinienne}
(Gauthier-Villars, Paris, 1921); V. A. Fock, {\it Theory of Space, Time and
Gravitation} (Pergamon, New York, 1959).

\bibitem{Hooft} G. 't Hooft, Nucl. Phys. B{\bf 72}, 461 (1974).

\bibitem{Gomis} J. Gomis and T. Mehen, hep-th/0005129.

\bibitem{Seiberg} N. Seiberg, L. Susskind, and N. Toumbas, JHEP 0006 (2000)
044, hep-th/0005015 v3.

\bibitem{Seiberg2} N. Seiberg, L. Susskind, and N. Toumbas, JHEP 0006
(2000) 021, hep-th/0005040.

\bibitem{Eliezer} D. A. Eliezer and R. P. Woodard, Nucl. Phys. B{\bf 325},
389 (1989); R. P. Woodard, hep-th/0006207.

\bibitem{Kamimura} J. Gomis, K. Kamimura and J. Llosa, hep-th/0006235.


\end{thebibliography}
\end{document}